\documentclass[sigplan,screen]{acmart}

\setcopyright{acmcopyright}
\copyrightyear{2023}
\acmYear{2023}
\acmDOI{XXXXXXX.XXXXXXX}

\acmConference[DAV '23]{First International Workshop on Deep Learning-aided Verification}{July 18, 2023}{Paris, France}

\acmPrice{15.00}
\acmISBN{978-1-4503-XXXX-X/18/06}

\usepackage{amsmath}
\usepackage{amsfonts}
\usepackage{xspace}
\usepackage{wrapfig}
\usepackage{multirow}
\usepackage{tikz}
\usepackage{caption}
\usepackage{subcaption}
\usepackage{nicefrac}
\usepackage{booktabs}
\usepackage{placeins}

\usepackage{pgfplots}
\pgfplotsset{width=\linewidth,height=4cm,compat=1.12}

\usepackage{standalone}
\standaloneconfig{mode=build}

\newcommand{\set}[1]{\left\{ #1 \right\}}
\newcommand{\card}[1]{\left| #1 \right|}

\newcommand{\up}{\textsc{UP}}
\newcommand{\coup}{\textsc{coUP}}

\newcommand{\nats}{{\mathbb N}}

\newcommand{\reals}{{\mathbb R}}

\newcommand{\arena}{\ensuremath{\mathcal A}}

\newcommand{\game}{\ensuremath{\mathcal G}}

\newcommand{\neigh}{\ensuremath{\mathcal N}}
\newcommand{\dgr}{\text{deg}}

\newcommand{\winreg}{W}

\newcommand{\err}{\mathrm{err}}

\newcommand{\col}{\Omega}

\newcommand{\pgsolver}{\texttt{PGSolver}\xspace}

\newcommand{\parity}[1]{\ensuremath{\textsc{Parity}(#1)}}

\begin{document}

\title{Predicting Winning Regions in Parity Games via Graph Neural Networks}
\author{Tobias Hecking}
\affiliation{%
  \institution{German Aerospace Center (DLR)\\Institute for Software Technology}
  \streetaddress{Linder H{\"o}he}
  \postcode{51147}
  \city{Cologne}
  \country{Germany}
}
\email{tobias.hecking@dlr.de}
\authornote{Authors are listed in alphabetical order and have contributed equally to this work}
\author{Swathy Muthukrishnan}
\affiliation{%
  \institution{University of Stuttgart}
  \city{Stuttgart}
  \country{Germany}
}
\email{swathy.muthukrishnan.24@gmail.com}
\authornotemark[1]
\author{Alexander Weinert}
\affiliation{%
  \institution{German Aerospace Center (DLR)\\Institute for Software Technology}
  \streetaddress{Linder H{\"o}he}
  \postcode{51147}
  \city{Cologne}
  \country{Germany}
}
\email{alexander.weinert@dlr.de}
\authornotemark[1]

\begin{abstract}
  Solving parity games is a major building block for numerous applications in reactive program verification and synthesis.
  While they can be solved efficiently in practice, no known approach has a polynomial worst-case runtime complexity.
  We present a incomplete polynomial-time approach to determining the winning regions of parity games via graph neural networks.

  Our evaluation on 900 randomly generated parity games shows that this approach is effective and efficient in practice.
  It correctly determines the winning regions of $\sim$60\% of the games in our data set and only incurs minor errors in the remaining ones.
  We believe that this approach can be extended to efficiently solve parity games as well.
\end{abstract}
  
\maketitle

\section{Introduction}
\label{sec:introduction}

Parity games are infinite arena-based games between Player~$0$ and Player~$1$ that capture all $\omega$-regular languages.
They are the canonical model for specifications of reactive systems, where Player~$0$ and Player~$1$ represent the system and its environment, respectively.
Solving them is a major building block for reactive program verification and synthesis.~\cite{Thomas2002,Fearnley2017}
Solving comprises determining the winning regions of both players as well as a winning strategy for both players.

The problem of solving parity games is known to be in $\up \cap \coup$~\cite{Jurdzinski1998} and to be solvable in quasi-polynomial time~\cite{CaludeJainKhoussainovEtAl2020}.
There exist parity game solvers that are efficient in practice~\cite{FriedmannLange2009,Dijk2018,StasioMuranoPrignanoEtAl2021}, all of which suffer from a non-polynomial worst-case runtime complexity.

Here, we trade completeness for efficiency in determining the winning regions.
Classically, parity games are solved by building a correct-by-construction strategy, which yields the winning regions as a by-product.
We instead train graph neural networks (GNNs)~\cite{BattagliaHamrickBapstEtAl2018} to determine the winning regions.
Applying the trained GNN to a parity game~$\game$ then yields a prediction of the winning region of~$\game$.

In a preliminary evaluation we constructed two GNNs and trained these on 2\thinspace100 randomly generated parity games.
We then predicted the winning region of another 900 randomly generated parity games using these GNNs.
This approach yields correct winning regions in 537 (433) cases, as well as regions that differ by one vertices from the correct result in 245 (279) cases.

\section{Preliminaries}
\label{sec:preliminaries}

We write~$\reals$ and~$\reals^{k,l}$ to denote the real numbers and the set of~$k\times l$-matrices over the reals, respectively.
Moreover, given a set~$S$, we write~$S^n$ to denote the $n$-ary cartesian power of~$S$.

\paragraph{Parity Games}
\label{sec:preliminaries:parity-games}

An arena~$\arena = (V, V_0, V_1, E)$ comprises a finite set of vertices~$V$, a partition~$(V_0, V_1)$ of~$V$, as well as a set of edges~$E \subseteq V \times V$ with $\forall v \in V.\, (\set{v} \times V) \cap E \neq \emptyset$.
We call~$V_i$ the vertices of Player~$i$.
A play~$\rho = v_0 v_1 v_2 \cdots$ of~$\arena$ is an infinite path through~$(V, E)$.
A parity game $(\arena, \parity{\col})$ comprises an arena~$\arena$ with vertex set~$V$ and a coloring~$\col\colon V \rightarrow \nats$.
The play~$\rho$ is winning for Player~$i$ if~$(\max\set{\col(v_i) \mid i \in \nats}) \mod 2 = i$.

A strategy~$\sigma \colon V_i \rightarrow V$ for Player~$i$ is a mapping with $\forall v \in V_i.\, (v, \sigma(v)) \in E$.
A play~$\rho = v_0 v_1 v_2 \cdots$ is consistent with~$\sigma$ if $\forall j \in \nats.\, v_j \in V_i \Rightarrow v_{j+1} = \sigma(v_j)$.
We say that Player~$i$ wins~$\game$ from vertex~$v \in V$ if she has a strategy~$\sigma$ such that all plays starting in~$v$ and consistent with~$\sigma$ are winning for her.
The winning region~$\winreg_i(\game)$ of Player~$i$ in~$\game$ is the set of all vertices from which Player~$i$ wins~$\game$.

\paragraph{Neural Network Layers}
\label{sec:preliminaries:neural-networks}

A~$k,l$-neural network layer is a function~$h\colon \reals^k \rightarrow \reals^l$.
In this work we use the linear layer $ h_{A,b}(x) = xA^T + b$, 
where~$A \in \reals^{l \times k}$ and~$b \in \reals^{1 \times l}$, the rectified linear unit (ReLU) layer
\[ h((x_1, \dots, x_k)^T) = (\max(0,x_1), \dots, \max(0, x_k))^T \enspace , \]
and the softmax layer
\[ h((x_1, \dots, x_k)^T) = (\nicefrac{e^{x_1}}{\Sigma_{1 \leq i \leq j}e^{x_i}}, \dots, \nicefrac{e^{x_n}}{\Sigma_{1 \leq i \leq j}e^{x_i}})^T \enspace . \]

\paragraph{Graph Neural Networks}
\label{sec:preliminaries:graph-neural-networks}

A $k$-attributed graph~$G = (V, E, X)$ comprises a graph~$(V, E)$ and a vertex-labeling~$X\colon V \rightarrow \reals^k$.
A graph neural network (GNN) in its general form maps a $k$-attributed input graph $G$ to an isomorphic~$l$-attributed output graph $G'$ by passing information through several layers.
Intuitively, each layer of an GNN can be seen as a function $h\colon \reals^k \rightarrow \reals^l$ that for each node aggregates the state of its neighbours and updates its state according to some rule.

Formally, for~$v \in V$ we define~$\neigh(v) = \set{v' \in V \mid (v, v') \in E}$ and~$\dgr(v) = \card{\neigh(v)}$.
Let~$\mathcal G_k$ denote the set of~$k$-attributed graphs.
A $k,l$-message-passing layer is a function~$h\colon \mathcal G_k \rightarrow (V \rightarrow \reals^l)$ where for all~$k$-attributed graphs $G = (V, E, X)$, $G' = (V, E, X')$ and all vertices~$v \in V$ we have
\begin{multline*}
\left(\forall v' \in \set{v} \cup \neigh(v).\, X(v') = X'(v') \right) \Rightarrow \\ h(G)(v) = h(G')(v) \enspace .
\end{multline*}
Intuitively, we require that~$h$ only updates the attributes of a vertex~$v$ based on the attributes of its neighbors.

In this work we evaluate two well-established message passing layers that differ in the way neighbourhood information about nodes is updated.
The first model uses a $k,l$-graph convolutional (GCN) layer, which was described by Kipf and Welling~\cite{KipfWelling2017}:
\[ h_{\mathbf W}((V, E, X)) \colon v \mapsto \mathbf{W} (\sum_{v' \in \set{v} \cup \neigh(v)}(\nicefrac{X(v')}{\sqrt{\dgr(v) \dgr(v')}})) \enspace , \]
where $\mathbf W$ is a $l \times k$-matrix.
The operator aggregates the attributes in the neighborhood of~$v$ and weights them by their degree.

In contrast, a $k,l$-graph attention layer (GAT) also aggregates the attributes in the neighborhood of~$v$, but weight these attributes by so-called attention weights:
\[ h_{\mathbf W, \alpha_{v, v'}}((V, E, X)) \colon v \mapsto \mathbf{W} (\sum_{v' \in \set{v} \cup \neigh(v)} \alpha_{v,v'} X(v') ) \enspace , \]
where~$\mathbf W$ is again a~$l \times k$-matrix and where~$\alpha_{v, v'}$ is the attention score for the pair~$(v, v')$.
We omit the definition of the~$\alpha_{v, v'}$ for brevity and refer the interested reader to work by Veli{\v{c}}kovi{\'c}, Cucurull, Casanova, et al.\, for details~\cite{VelivckovicEtAl2018}.

\section{Method}
\label{sec:method}

We aim to compare the performance of the GCN layer and the GAT layer in predicting winning regions.
Since both layers take attributed graphs as input and produce node labelings, we need to encode the problem of determining winning regions as a vertex-labeling problem.
In particular, we need to encode the color as well as the owner of a vertex.
Experience shows that the na\"ive encoding~$x(v) = (\col(v), p)$ with~$p = 0$ if~$v \in V_0$ and~$p = 1$ otherwise and using a~$2,1$-message passing layer does not yield satisfactory results.

Instead, experience and preliminary evaluation leads us to the following architecture:
Given a parity game~$\game = (\arena, \parity{\col})$ with~$\arena = (V, V_0, V_1, E)$ we define~$G = (V, E, X_0)$ with~$X_0(v) = (\col(v), x_0, x_1)$ where~$x_i = 1$ if~$v \in V_i$ and~$x_i = 0$ otherwise.
We then pass this~$2$-attributed graph to a stack of~$10$ message-passing layers, each of which yields a~$256$-attributed graph, i.e., for each vertex we obtain a vector of size~$256$.
For each vertex, we pass this vector to a stack of neural network layers comprising a $256,256$ linear layer, a $256,256$ ReLU layer, a $256,2$ linear layer, and a $2,2$ softmax layer.
Thus, for each vertex~$v \in V$ we obtain a vector~$(x_0, x_1)$.
We interpret this vector such that the architecture predicts~$v \in \winreg_0(\game)$ if~$x_0 > x_1$ and~$v \in \winreg_1(\game)$ otherwise.

Recall that most layers in our architecture are parameterized.
Hence, it remains to determine optimal parameters for these layers for the task of predicting winning regions.
As with most machine learning tasks it is infeasible to determine these parameters analytically.
Instead, we approximate them as follows:
Let~$\set{\game_1,\dots,\game_n}$ be a set of parity games with arenas~$\arena_1,\dots,\arena_n$ and let~$\winreg^j_i(\game)$ be the winning region of Player~$i$ in game~$j$.
For each arena~$\arena_j = (V, V_0, V_1, E)$ we construct the~$k$-attributed graph $G_j=(V, E, X)$ with~$X(v) = (\col(v), p_0, p_1)$ with~$p_i = 1$ if $v \in V_i$ and $p_i = 0$ otherwise.
Moreover, for each~$\game_j$ we define the ``target'' vertex attributes~$X'_j(v) = (w_0, w_1)$, where~$w_i = 1$ if~$v \in \winreg_i(\game_j)$ and~$w_i = 0$ otherwise.
Using these pairs of inputs and desired outputs we approximate the optimal parameters of the architectures using the Adam algorithm~\cite{KingmaBa2015}.
For technical reasons, we additionally use an so-called Dropout layer~\cite{HintonSrivastavaKrizhevskyEtAl2012} after the ReLU-layer.
This improves the performance of the optimization, but does not change the function computed by the neural networks.

Having obtained approximations of the optimal parameters on the input data described above, we can apply either model to a previously unseen game~$\game = (\arena, \parity{\col})$ and obtain vertex-attributes~$X'_V\colon V \rightarrow \reals^2_+$.
Given these attributes, we define the predicted winning regions $\winreg^p_0(X'_V) = \set{v \in V \mid X'_V(v) = (w_0, w_1), w_0 > w_1}$ and $\winreg^p_1 = V \setminus \winreg^p_0(X'_V)$.

The application of the model on~$\game$ comprises the application of the ten message-passing layers as well as the application of each of the neural network layers for each vertex.
Applying a single message-passing layer to takes constant time per vertex, i.e., linear time in the number of vertices of the arena.
Similarly, the application of each neural network layer takes constant time.
Hence, we are able to obtain the predicted winning regions in linear time in the number of vertices of the arena of~$\game$.

\section{Evaluation}
\label{sec:evaluation}

\begin{table}
\centering
\caption{Results of the evaluation. Acc. denotes the share of correctly identified vertices of the total number of vertices in the training set, while $n_\err = 0$, $n_\err = 1$, $n_\err \geq 2$ denote the number of games with zero, one, and at least two misclassified vertices in the test set, respectively.}
\label{tab:results1}
\begin{tabular}{lcccc} \toprule
  \textbf{Model} & Acc. & $n_\err = 0$ & $n_\err = 1$ & $n_\err \geq 2$ \\ \midrule
GCN & 0.60 & 537 & 245 & 118 \\
GAT & 0.98 & 433 & 279 & 188 \\ \bottomrule
\end{tabular}
\end{table}

\begin{figure}
  \centering
  \includegraphics[width=\linewidth]{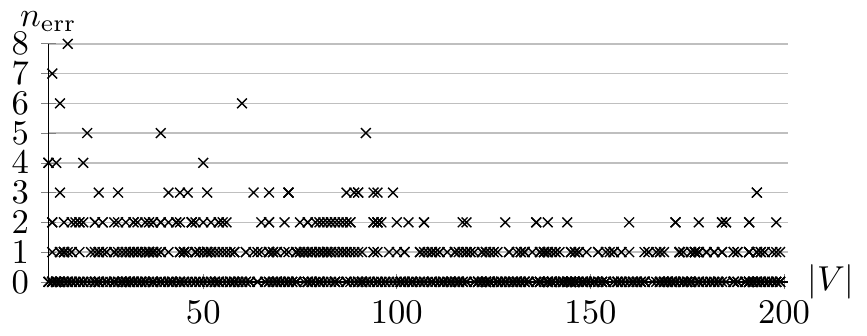}
  \caption{Misclassified vertices in relation to the size of the arena for GCN.}
  \label{fig:correlation-size-regions:gcn}
\end{figure}

\begin{figure}
  \centering
  \includegraphics[width=\linewidth]{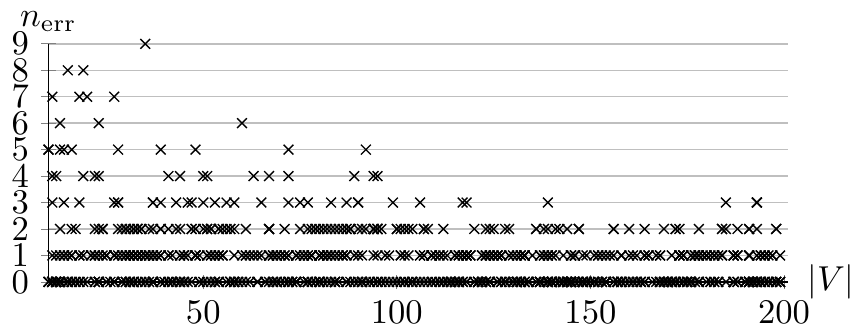}
  \caption{Misclassified vertices in relation to the size of the arena for GAT.}
  \label{fig:correlation-size-regions:gat}
\end{figure}

To evaluate our approach we have developed a prototypical implementation~\cite{HeckingWeinert2022a} and generated 3\thinspace000 parity games and their solutions~\cite{HeckingWeinert2022b} using \pgsolver~\cite{FriedmannLange2014,FriedmannLange2009}.
We sampled the number of vertices uniformly from $\set{10,\dots,200}$.
In each game with~$n$ vertices, for each vertex~$v$, we sampled the number of outgoing edges and~$\col(v)$ from $\set{\frac{n}{100},\dots,\frac{n}{2}}$ and $\set{1,\dots,n}$, respectively.
We trained both models on 2\thinspace100 of these games (the training set) and evaluated their performance on the remaining 900 games (the test set).
We show the results in Table~\ref{tab:results1}.

During training we observed that both models converge very fast, namely after processing the first~$50$ games.
Our results show that GCN cannot characterize vertices in games it has not seen during training, it assigns vertices to the winning regions correctly for around 60\% of previously unseen games.
GAT, in contrast, shows better performance in this aspect.
However, when it comes to solving entire games correctly, it only correctly predicts winning regions entirely for around half of previously unseen games.
This in the first view contradicting results can be explained by examining the relation between the size of games and the number of incorrectly identified vertices in Figure~\ref{fig:correlation-size-regions:gcn} and Figure~\ref{fig:correlation-size-regions:gat}.
GAT fails more often in small games compared to GCN but performes better on larger ones.
Consequently, the overall accuracy on vertex level is better for GAT while solving less games entirely correctly than GCN.
We leave further research on this, however, to future research.

\section{Conclusion and Future Work}
\label{sec:conclusion}

We have described a novel method for predicting the winning regions in parity games and we have evaluated a prototypical implementation of this method using randomly generated parity games.
This evaluation showed that our method correctly identifies $\sim$60\% of the winning regions for Player~$0$ in our data set.
We strongly believe that these results indicate the feasibility of our approach to determining the winner of parity games in practice.
There are, however, open questions left for future research.

One direction of future work is to apply GNNs trained on smaller games to larger games.
The correlation between game size and quality of prediction shown in Figure~\ref{fig:correlation-size-regions:gcn} and in Figure~\ref{fig:correlation-size-regions:gat} indicates that we may be able to retain the prediction quality on larger games as well.
If this is the case, it might be possible to use games of sizes that can be handled by exact solvers to create training data, which can then be used to train graph neural networks that determine winning regions of games of sizes intractable by classical methods. 

Another avenue for future research concerns our architecture.
In particular, the number of message-passing layers, hidden features, and a possible relationship with game size need further investigations.  

Moreover, we are currently only predicting the winner of a parity game for a given vertex, but obtain no witness of this prediction.
While the GNNs can easily be extended to produce edge attributes as well as node attributes, preliminary experiments show that the former cannot be used to construct winning strategies for the player predicted to win from a given vertex.
In future work, we will look into predicting winning strategies as well as winning regions.

We also aim to evaluate our approach in a more realistic setting.
This includes, e.g., solving parity games to construct controllers satisfying specifications given in LTL~\cite{PnueliRosner1989,LuttenbergerMeyerSickert2019}.

Finally, in our current approach we only use a binary payoff function:
Either Player~$0$ wins from a vertex, or she does not.
In recent years more granular metrics for strategies have emerged.
These include, e.g., evaluating the robustness of strategies against external disturbances~\cite{NeiderWeinertZimmermann2019} or the number of steps until an odd number is followed by a larger even number~\cite{WeinertZimmermann2016,ScheweWeinertZimmermann2018}.
In future work, we aim to use graph neural networks to evaluate vertices according to these metrics.

\bibliographystyle{ACM-Reference-Format}
\bibliography{overlay22-learning-parity-game-strategies}

\end{document}